\documentclass[12pt,superscriptaddress,aps,prd,preprint,showpacs]{revtex4}
\usepackage{graphicx}
\usepackage{amsfonts}
\usepackage{slashed}
\usepackage[utf8x]{inputenc}
\usepackage{amsmath}
\usepackage{hyperref}

\begin{document}

\title{Non-Abelian Carroll–Field–Jackiw term in a Rarita-Schwinger model}

\author{M. Gomes}
\affiliation{Instituto de F\'\i sica, Universidade de S\~ao Paulo\\
Caixa Postal 66318, 05315-970, S\~ao Paulo, SP, Brazil}
\email{mgomes@if.usp.br}

\author{J. G. Lima}
\affiliation{Instituto de F\'\i sica, Universidade Federal de Alagoas,\\ 57072-900, Macei\'o, Alagoas, Brazil}
\email{grimario.lima,tmariz@fis.ufal.br}

\author{T. Mariz}
\affiliation{Instituto de F\'\i sica, Universidade Federal de Alagoas,\\ 57072-900, Macei\'o, Alagoas, Brazil}
\email{grimario.lima,tmariz@fis.ufal.br}

\author{J. R. Nascimento}
\affiliation{Departamento de F\'\i sica, Universidade Federal da Para\'\i ba,\\
 Caixa Postal 5008, 58051-970, Jo\~ao Pessoa, Para\'\i ba, Brazil}
\email{jroberto,petrov@fisica.ufpb.br}

\author{A. Yu. Petrov}
\affiliation{Departamento de F\'\i sica, Universidade Federal da Para\'\i ba,\\
 Caixa Postal 5008, 58051-970, Jo\~ao Pessoa, Para\'\i ba, Brazil}
\email{jroberto,petrov@fisica.ufpb.br}

\begin{abstract}
In this paper, we demonstrate the possibility of generating  a  non-Abelian Carroll–Field–Jackiw (CFJ) term in the theory of a non-Abelian gauge field coupled to a spin-3/2 field in the presence of the constant axial vector field. Applying two regularization schemes, we prove that this term is finite and ambiguous, particularly vanishing within the 't Hooft-Veltman scheme.
\end{abstract}


\maketitle

\section{Introduction}

The study of Lorentz-breaking extensions for various field theory models gave origin to the Lorentz-violating standard model extension (LV SME) \cite{ColKost1,ColKost2}. Within this theory, all possible minimal LV couplings of scalar, spinor, and gauge fields (both Abelian and non-Abelian ones) are considered. A further development of the LV SME allowed the inclusion of gravity and the introduction of LV gravity-dependent terms \cite{KosGra}. In numerous papers, various aspects of new LV models were studied, such as dispersion relations, exact solutions, quantum corrections, modifications of known spacetime metrics, and so on. Nevertheless, the list of models presented in \cite{ColKost1,ColKost2,KosGra} is actually only a partial one. First of all, nonminimal terms involving higher-dimensional operators, and, in particular, higher derivative terms, can be introduced as well. A list of such terms with dimensions up to 6 is presented for non-gravitational theories in \cite{KosLi1}, and for the presence of gravity -- in \cite{KosLi2}. Second, it is interesting to construct LV models involving other fields. Certainly, the very natural candidate to be introduced within LV theories is the spin-3/2 Rarita-Schwinger (RS) field \cite{RS}, known for applications within the supergravity context (see, e.g., \cite{SGRS}) and studied within the phenomenological context as well,  see e.g.. \cite{Delgado} for scattering of spin-3/2 particles, and \cite{Delta} and references therein for study of $\Delta$ baryons representing themselves as observed examples of spin-3/2 particles. Some tree-level aspects of the theory of the RS field, including the derivation of propagators, are discussed in \cite{Pascal,Pill}. In our previous paper \cite{prev}, we coupled the RS field to the Abelian vector  by adding the LV term proportional to $\slashed{b}\gamma_5$  which is a natural generalization of the term $\bar{\psi}\slashed{b}\gamma_5\psi$ treated within the standard minimal LV QED. Now, we generalize the results obtained there to the non-Abelian case, reproducing the non-Abelian Carroll–Field–Jackiw (CFJ) term discussed in great detail in \cite{ColMac} and shown in \cite{ourYM} to arise as a quantum correction in a theory where non-Abelian gauge field is coupled to usual fermions; our   result turns out to be finite but ambiguous.

The structure of the paper looks as follows. In section 2, we write down our Lagrangian and the corresponding Feynman rules, i.e., the propagator, the insertion, and the vertex. Section 3 calculates the one-loop contribution to the non-Abelian CFJ term through the two-point and three-point functions. Finally, in Summary  (section 4), we discuss our results.

\section{Lagrangian and Feynman Rules}

The starting point of our study is the Lagrangian of the spin-3/2 field further referred as RS field, coupled to the Yang-Mills field with the inclusion of a Lorentz-breaking term proportional to the constant axial vector $b^\mu$, given by
\begin{equation}
{\cal L}=\bar{\psi}_{\mu}\frac{i}{2}\{\sigma^{\mu\nu},(i\slashed{\partial}+g\slashed{A^a}T^a -m -\slashed{b}\gamma_5)\}\psi_{\nu},
\end{equation}
where $\sigma^{\mu\nu}=\frac{i}{2}[\gamma^\mu,\gamma^\nu]$ is the Dirac sigma matrix, and $A_\mu=A^a_\mu T^a$ is the Lie-algebra valued Yang-Mills vector field, with $T^a$ being the generators of some Lie group algebra satisfying the relations $tr(T^aT^b)=\delta^{ab}$ and $[T^a,T^b]=if^{abc}T^c$. We note that the Lagrangian of the form $\bar{\psi}_{\mu}\frac{i}{2}\{\sigma^{\mu\nu},(i\slashed{\partial}+g\slashed{A^a}T^a -m)\}\psi_{\nu}$ we use here, which is slightly different from that one used in our previous paper \cite{prev}, seems to be more appropriate for introducing Lorentz symmetry breaking, being, in a certain sense, the analog of the generalized LV Lagrangian for the spinor QED \cite{KosPic}.

The above Lagrangian can be rewritten as
\begin{eqnarray}
{\cal L}&=&\bar{\psi}_{\mu}\left((i\slashed{D}-m)g^{\mu\nu} 
-i(\gamma^{\mu}D^{\nu}+\gamma^{\nu}D^{\mu}) +i\gamma^{\mu}\slashed{D}\gamma^{\nu} +m\gamma^{\mu}\gamma^{\nu} \right.\nonumber\\
&&\left.+\slashed{b}\gamma_5 g^{\mu\nu} -(\gamma^{\mu}b^{\nu}+\gamma^{\nu}b^{\mu})\gamma_5 +\gamma^{\mu}\slashed{b}\gamma^{\nu}\gamma_5\right)\psi_{\nu},
\end{eqnarray}
where $D_\mu=\partial_\mu- ig A^a_\mu T^a$ is the covariant derivative. It is easy to observe that the above expression is the most usual form of the free spin-$3/2$ Lagrangian corresponding to the choice $A=-1$, for more details, see Ref.~\cite{Pill,prev}.

Let us now consider the Feynman rules. For the spin-$3/2$ field propagator in $D$ dimensions, we have (cf. \cite{Pill}, with $A=-1$):
\begin{equation}
i G^{\mu\nu}(p)=i \frac{\slashed{p}+m}{p^2-m^2}\left(g^{\mu\nu}-\frac{1}{D-1}\gamma^{\mu}\gamma^{\nu}-\frac{D-2}{D-1}\frac{p^{\mu}p^{\nu}}{m^2}-\frac{1}{D-1}\frac{\gamma^{\mu}p^{\nu}-\gamma^{\nu}p^{\mu}}{m}\right).
\end{equation}
We note that the above expression depends essentially on the spacetime dimension. This form is appropriate within the dimensional regularization framework we use in the paper. The coefficient for Lorentz and CPT violation $b_\mu$ leads to an insertion in the spin-3/2 field propagator, given by
\begin{equation}
-i (g^{\mu\nu}\slashed{b}\gamma_5 -(\gamma^{\mu}b^{\nu}+\gamma^{\nu}b^{\mu})\gamma_5 +\gamma^{\mu}\slashed{b}\gamma^{\nu}\gamma_5) = -i b_{\lambda}\gamma^{\mu\lambda\nu}\gamma_5,
\end{equation}
where $\gamma^{\mu\lambda\nu}=g^{\mu\nu}\gamma^\lambda -g^{\mu\lambda}\gamma^\nu -g^{\nu\lambda}\gamma^\mu +\gamma^\mu\gamma^\lambda\gamma^\nu$. For the RS-photon vertex (that we refer to as the minimal one), we write
\begin{equation}
ig \left(g^{\mu\nu}\gamma^\lambda -(\gamma^{\mu}g^{\nu\lambda}+\gamma^{\nu}g^{\mu\lambda}) +\gamma^\mu\gamma^\lambda\gamma^\nu\right)T^a = ig \gamma^{\mu\lambda\nu}T^a.
\end{equation}
These vertices will be used for constructing the Feynman diagrams.

\section{Two-point and three-point functions}

\begin{figure}[htbp]
  \includegraphics[width=4cm]{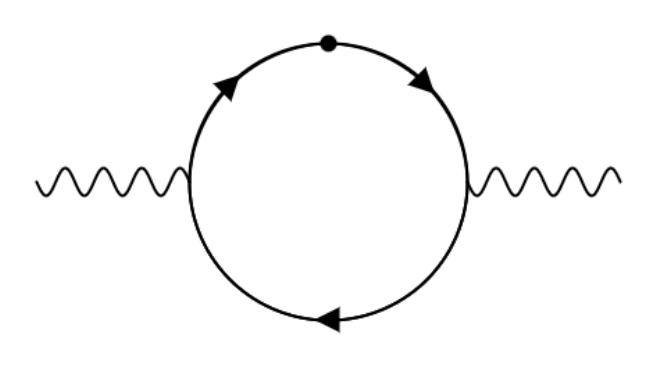}  
  \includegraphics[width=4cm]{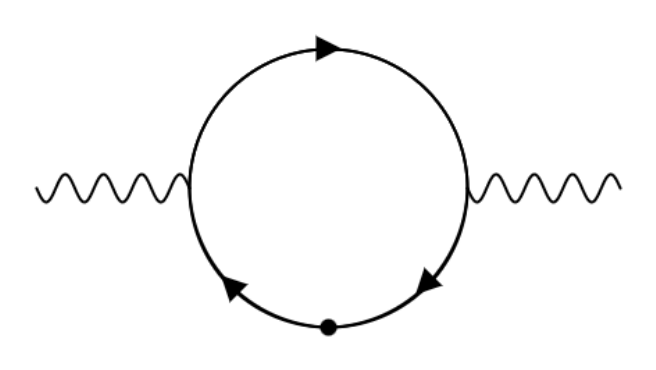}    
\caption{CFJ contributions to the two-point function of the vector field.}
\label{fig:1}       
\end{figure}

Let us write down the contributions to the two and three-point functions of the gauge field. The two-point one, in which the graphs are depicted in Fig.~\ref{fig:1}, is similar to that in the Abelian case \cite{prev} but involves an additional trace related to Lie algebra generators.  

The corresponding action, with one insertion of the coefficient $b_\mu$ in the propagator $G^{\mu\nu}(p)$, is given by
\begin{equation}
S^{(2)}_{CFJ} = \frac{i}{2} \int \frac{d^4p}{(2\pi)^4} (\Pi_1^{ab\lambda\tau}+\Pi_2^{ab\lambda\tau})A^a_{\lambda}(-p)A^b_{\tau}(p)
\end{equation}
with
\begin{subequations}
\begin{eqnarray}
\label{twomin}
\Pi_1^{ab\lambda\tau} &=& -g^2{\rm tr}(T^aT^b)\,{\rm tr}\int\frac{d^4k}{(2\pi)^4}
\gamma^{\mu\lambda\nu} G_{\nu\alpha}(k) b_\kappa\gamma^{\alpha\kappa\beta}\gamma_5 G_{\beta\rho}(k) \gamma^{\rho\tau\sigma} G_{\sigma\mu}(k+p),\\
\Pi_2^{ab\lambda\tau} &=& -g^2{\rm tr}(T^aT^b)\,{\rm tr}\int\frac{d^4k}{(2\pi)^4}
\gamma^{\mu\lambda\nu} G_{\nu\rho}(k) \gamma^{\rho\tau\sigma} G_{\sigma\alpha}(k+p) b_\kappa\gamma^{\alpha\kappa\beta}\gamma_5 G_{\beta\mu}(k+p).
\end{eqnarray}
\end{subequations}
So, by expanding $G^{\mu\nu}(k+p)$ up to the first order in external $p$, we can write $G^{\mu\nu}(k+p)=G^{\mu\nu}(k)+G^{\mu\nu}(k,p)+{\cal O}(p^2)$, where
\begin{eqnarray}
G^{\mu\nu}(k,p) &=& -\frac{1}{\slashed{k}-m}\slashed{p}\frac{1}{\slashed{k}-m}(g^{\mu\nu}-\frac{1}{D-1}\gamma^{\mu}\gamma^{\nu}-\frac{D-2}{D-1}\frac{k^{\mu}k^{\nu}}{m^2}-\frac{1}{D-1}\frac{\gamma^{\mu}k^{\nu}-\gamma^{\nu}k^{\mu}}{m})\nonumber\\ 
&& -\frac{1}{\slashed{k}-m}
\left(\frac{D-2}{D-1}\frac{k^{\mu}p^{\nu}+k^{\nu}p^{\mu}}{m^2}+\frac{1}{D-1}\frac{\gamma^{\mu}p^{\nu}-\gamma^{\nu}p^{\mu}}{m}\right).
\end{eqnarray}
In a total analogy with the Abelian case \cite{prev}, we can write
\begin{subequations}
\label{subtwomin1}
\begin{eqnarray}
\label{twomin1}
\Pi_{1a}^{ab\lambda\tau} &=& -g^2{\rm tr}(T^aT^b)\,{\rm tr}\int\frac{d^4k}{(2\pi)^4}
\gamma^{\mu\lambda\nu} G_{\nu\alpha}(k) b_\kappa\gamma^{\alpha\kappa\beta}\gamma_5 G_{\beta\rho}(k) \gamma^{\rho\tau\sigma} G_{\sigma\mu}(k,p),\\
\Pi_{2a}^{ab\lambda\tau} &=& -g^2{\rm tr}(T^aT^b)\,{\rm tr}\int\frac{d^4k}{(2\pi)^4}
\gamma^{\mu\lambda\nu} G_{\nu\rho}(k) \gamma^{\rho\tau\sigma} G_{\sigma\alpha}(k,p) b_\kappa\gamma^{\alpha\kappa\beta}\gamma_5 G_{\beta\mu}(k),\\
\Pi_{2b}^{ab\lambda\tau} &=& -g^2{\rm tr}(T^aT^b)\,{\rm tr}\int\frac{d^4k}{(2\pi)^4}
\gamma^{\mu\lambda\nu} G_{\nu\rho}(k) \gamma^{\rho\tau\sigma} G_{\sigma\alpha}(k) b_\kappa\gamma^{\alpha\kappa\beta}\gamma_5 G_{\beta\mu}(k,p).
\end{eqnarray}
\end{subequations}

Notice that these expressions involve the computations of the trace of  just one $\gamma_5$ matrix times a product of usual Dirac $\gamma$'s matrices.  Requiring the Dirac matrices in an arbitrary space-time dimension $D$ in such a manner  that the anticommutation relation $\{\gamma_\mu,\gamma_5\} =0$ is valid at $D=4-\varepsilon$ as well, we move the $\gamma_5$ matrix to the utmost right position and  compute the traces as in four dimensions (${\rm tr} \,\, \gamma_\kappa\gamma_\lambda\gamma_\mu\gamma_\nu\gamma_5  =4i \epsilon_{\kappa\lambda\mu\nu}$, for example).  We then extent the calculation to $D$ dimensions 
so that $d^4k/(2\pi)^4$ goes to $\mu^{4-D}d^Dk/(2\pi)^D$, with $\mu$ being an arbitrary scale parameter with the mass dimension 1 and  $g_{\mu\nu}g^{\mu\nu}=D$ (actually, this procedure is the dimensional reduction \cite{Siegel}). 
Thus we will obtain
\begin{subequations}
\begin{eqnarray}
\label{twomin1a00}
\Pi_{1a}^{ab\lambda\tau} &=& \mu^{4-D}\int\frac{d^Dk}{(2\pi)^D}\frac{4ig^2 \delta^{ab}\epsilon^{\mu\nu\lambda\tau}b_\mu p_\nu}{(D-1)^3 D m^2} \left[\frac{2 (D (3 (D-11) D+152)-350) D+696}{k^2-m^2}\right. \nonumber\\
&& -\frac{(D ((((D-9) D+9) D+89) D+86)-992) m^2}{(k^2-m^2)^2} \nonumber\\
&& \left.-\frac{2 (D (D (((D-9) D+11) D+85)-116)-260) m^4}{(k^2-m^2)^3}\right],\\
\Pi_{2a}^{ab\lambda\tau} &=& -\mu^{4-D}\int\frac{d^Dk}{(2\pi)^D}\frac{8ig^2 \delta^{ab}\epsilon^{\mu\nu\lambda\tau}b_\mu p_\nu}{(D-1)^3 D m^2} \left[\frac{2 (D (((D-7) D-3) D+83)-110)}{k^2-m^2}\right. \nonumber\\
&& +\frac{(D ((((D-1) D-47) D+185) D+86)-992) m^2}{2 (k^2-m^2)^2} \nonumber\\
&& \left.+\frac{(D (D (((D-9) D+11) D+85)-116)-260) m^4}{(k^2-m^2)^3}\right],\\
\Pi_{2b}^{ab\lambda\tau} &=& \mu^{4-D}\int\frac{d^Dk}{(2\pi)^D}\frac{4ig^2 \delta^{ab}\epsilon^{\mu\nu\lambda\tau}b_\mu p_\nu}{(D-1)^2 D m^2} \left[\frac{2 (((D-4) D+2) D+4)}{k^2-m^2}\right. \nonumber\\
&& +\frac{(D (((D-12) D+37) D+10)-144) m^2}{(k^2-m^2)^2} \nonumber\\
&& \left.-\frac{4 (((D-8) D+9) D+38) m^4}{(k^2-m^2)^3}\right].
\end{eqnarray}
\end{subequations}
Actually, the only difference from the Abelian case consists of the trace factor. Finally, by calculating the integrals in $D$ dimensions, the results are
\begin{subequations}
\begin{eqnarray}
\label{twomin1a}
\Pi_{1a}^{ab\lambda\tau} &=& -g^2\frac{2^{2-D} \pi^{-\frac{D}{2}} \mu^{4-D} m^{D-4}}{(D-2) (D-1)^3 D} \Gamma\left(3-\frac{D}{2}\right) \delta^{ab}\epsilon^{\mu\nu\lambda\tau}b_\mu p_\nu \nonumber\\
&& \times(D (D (D (D ((D-9) D+15)+37)+16)-220)+224),\\
\Pi_{2a}^{ab\lambda\tau} &=& -g^2\frac{2^{2-D} \pi^{-\frac{D}{2}} \mu^{4-D} m^{D-4}}{(D-2) (D-1)^3 D}\Gamma\left(3-\frac{D}{2}\right) \delta^{ab}\epsilon^{\mu\nu\lambda\tau}b_\mu p_\nu \nonumber\\
&& \times((D-2) (D-1) D (D ((D-6) D+11)+42)-32),\\
\Pi_{2b}^{ab\lambda\tau} &=& 0.
\end{eqnarray}
\end{subequations}
Then, in $D=4$, we have
\begin{subequations}
\begin{eqnarray}
\label{twomin1aa}
\Pi_{1a}^{ab\lambda\tau} &=& -g^2\delta^{ab}\frac{43}{54\pi^2}\epsilon^{\mu\nu\lambda\tau}b_\mu p_\nu,\\
\Pi_{2a}^{ab\lambda\tau} &=& -g^2\delta^{ab}\frac{79}{54\pi^2}\epsilon^{\mu\nu\lambda\tau}b_\mu p_\nu,\\
\Pi_{2b}^{ab\lambda\tau} &=& 0,
\end{eqnarray}
\end{subequations}
which are finite despite the integrals being badly (sextically) divergent by power counting. Adding these results, we get
\begin{eqnarray}
\label{twomin1aaa}
\Pi^{ab\lambda\tau} &=& \Pi_{1a}^{ab\lambda\tau} +\Pi_{2a}^{ab\lambda\tau} +\Pi_{2b}^{ab\lambda\tau} \nonumber\\
&=& -g^2\delta^{ab}\frac{61}{27\pi^2}\epsilon^{\mu\nu\lambda\tau}b_\mu p_\nu.
\end{eqnarray}

\begin{figure}[htbp]
  \includegraphics[width=4cm]{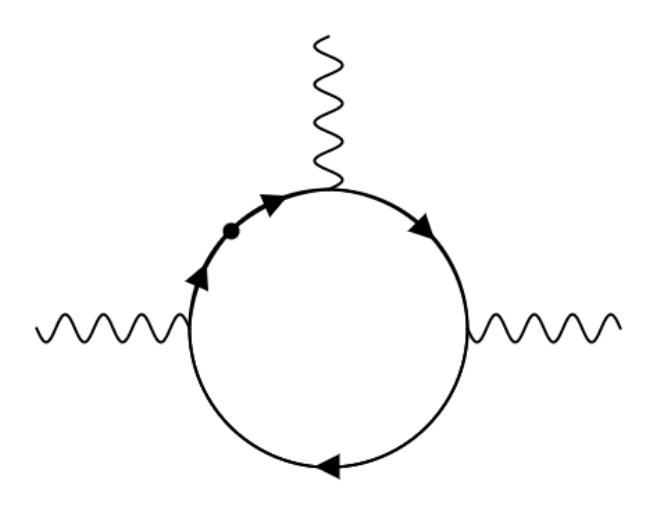} 
  \includegraphics[width=4cm]{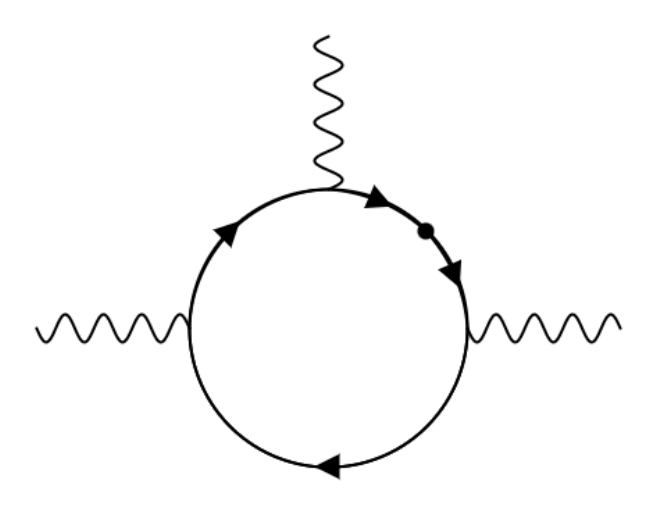}
  \includegraphics[width=4cm]{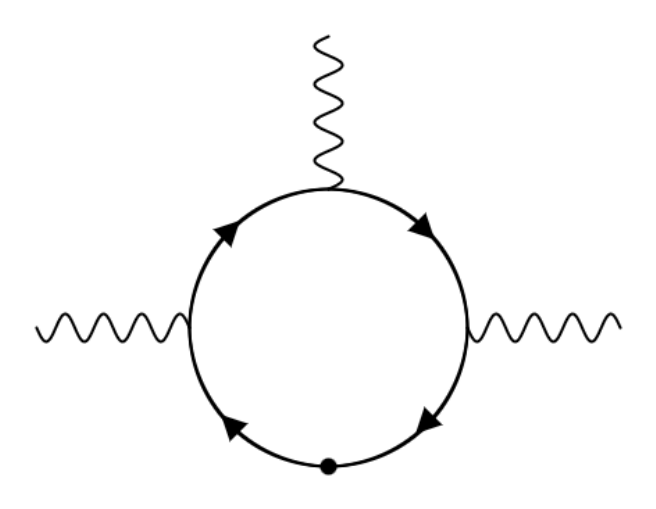}    
\caption{CFJ contributions to the three-point function of the vector field.}
\label{fig:2}       
\end{figure}

The next step is calculating the three-point function, whose graphs are depicted in Fig.~\ref{fig:2}. 

Since we are interested only in zero-order contributions to the three-point function, we can suppress the $p$ dependence of propagators. Therefore, in this case, we have the action 
\begin{equation}
S^{(3)}_{CFJ} = \frac{i}{3} \int \frac{d^4p_1}{(2\pi)^4} \int \frac{d^4p_2}{(2\pi)^4} (\Pi_1^{abc\lambda\tau\epsilon}+\Pi_2^{abc\lambda\tau\epsilon}+\Pi_3^{abc\lambda\tau\epsilon})A^a_{\lambda}(-p_1-p_2)A^b_{\tau}(p_1)A^c_{\epsilon}(p_2),
\end{equation}
with
\begin{subequations}
\label{subthreemin}
\begin{eqnarray}
\label{threemin}
\Pi_1^{abc\lambda\tau\epsilon} &=& g^3{\rm tr}(T^aT^bT^c)\,{\rm tr}\int\frac{d^4k}{(2\pi)^4} \gamma^{\mu\lambda\nu} G_{\nu\alpha}(k) b_\kappa\gamma^{\alpha\kappa\beta}\gamma_5 G_{\beta\rho}(k)
\gamma^{\rho\tau\sigma} G_{\sigma\delta}(k) \gamma^{\delta\epsilon\eta} G_{\eta\mu}(k), \hspace{0.7cm}\\
\Pi_2^{abc\lambda\tau\epsilon} &=& g^3{\rm tr}(T^aT^bT^c)\,{\rm tr}\int\frac{d^4k}{(2\pi)^4}
\gamma^{\mu\lambda\nu} G_{\nu\rho}(k) \gamma^{\rho\tau\sigma} G_{\sigma\alpha}(k) b_\kappa\gamma^{\alpha\kappa\beta}\gamma_5 G_{\beta\delta}(k) \gamma^{\delta\epsilon\eta} G_{\eta\mu}(k),\\
\Pi_3^{abc\lambda\tau\epsilon} &=& g^3{\rm tr}(T^aT^bT^c)\,{\rm tr}\int\frac{d^4k}{(2\pi)^4}
\gamma^{\mu\lambda\nu} G_{\nu\rho}(k) \gamma^{\rho\tau\sigma} G_{\sigma\alpha}(k)
\gamma^{\delta\epsilon\eta} G_{\eta\alpha}(k) b_\kappa\gamma^{\alpha\kappa\beta}\gamma_5 G_{\beta\mu}(k).
\end{eqnarray}
\end{subequations}

These expressions will be calculated following the same route chosen above to proceed with the tensors (\ref{subtwomin1}). After computing the trace, we obtain
\begin{subequations}
\begin{eqnarray}
\label{threemin1}
\Pi_1^{abc\lambda\tau\epsilon} &=& \mu^{4-D}\int\frac{d^Dk}{(2\pi)^D}\frac{4ig^3 f^{abc}\epsilon^{\mu\lambda\tau\epsilon}b_\mu}{(D-1)^3 D m^2} \left[\frac{2 ((D ((2 D-35) D+257)-804) D+868)}{k^2-m^2}\right. \nonumber\\
&& -\frac{(D ((D ((D-7) D+17)-109) D+834)-1840) m^2}{(k^2-m^2)^2} \nonumber\\
&& \left.-\frac{2 (D+1) \left(D \left(D (D-5)^2+52\right)-212\right) m^4}{(k^2-m^2)^3}\right],\\
\Pi_2^{abc\lambda\tau\epsilon} &=& \mu^{4-D}\int\frac{d^Dk}{(2\pi)^D}\frac{4ig^3 f^{abc}\epsilon^{\mu\lambda\tau\epsilon}b_\mu}{(D-1)^2 D m^2} \left[\frac{2 (D-2) ((2 D-7) D+2)}{k^2-m^2}\right. \nonumber\\
&& +\frac{(D (((D-10) D+23) D+14)-64) m^2}{(k^2-m^2)^2} \nonumber\\
&& \left.-\frac{4 (D+1) ((D-7) D+14) m^4}{(k^2-m^2)^3}\right], \\
\Pi_3^{abc\lambda\tau\epsilon} &=& -\mu^{4-D}\int\frac{d^Dk}{(2\pi)^D}\frac{8ig^3 f^{abc}\epsilon^{\mu\lambda\tau\epsilon}b_\mu}{(D-1)^3 D m^2} \left[\frac{D ((17 D-193) D+684)-772}{k^2-m^2}\right. \nonumber\\
&& +\frac{(D ((((D-3) D-19) D+19) D+594)-1648) m^2}{2(k^2-m^2)^2} \nonumber\\
&& \left.+\frac{\left(D \left(D (D-5)^2+52\right)-212\right) (D+1) m^4}{(k^2-m^2)^3}\right].
\end{eqnarray}
\end{subequations}
Now, performing the integration, we arrive at the following results:
\begin{subequations}
\begin{eqnarray}
\label{threemin10}
\Pi_1^{abc\lambda\tau\epsilon} &=& -g^3\frac{2^{2-D} \pi^{-\frac{D}{2}} \mu^{4-D} m^{D-4}}{(D-2) (D-1)^3 D}\Gamma\left(3-\frac{D}{2}\right) f^{abc}\epsilon^{\mu\lambda\tau\epsilon}b_\mu \nonumber\\
&& \times((D-1) D (D+3) (D ((D-11) D+48)-76)+320),\\
\Pi_2^{abc\lambda\tau\epsilon} &=& 0, \\
\Pi_3^{abc\lambda\tau\epsilon} &=& -g^3\frac{2^{2-D} \pi^{-\frac{D}{2}} \mu^{4-D} m^{D-4}}{(D-2) (D-1)^3 D} \Gamma\left(3-\frac{D}{2}\right) f^{abc}\epsilon^{\mu\lambda\tau\epsilon}b_\mu \nonumber\\
&& \times(D+2)(D (D (D ((D-11) D+53)-93)-14)+160).
\end{eqnarray}
\end{subequations}
Therefore, in $D=4$, we get
\begin{subequations}
\begin{eqnarray}
\label{threemin1a}
\Pi_1^{abc\lambda\tau\epsilon} &=& -g^3f^{abc}\frac{41}{54\pi^2} \epsilon^{\mu\lambda\tau\epsilon}b_\mu,\\
\Pi_2^{abc\lambda\tau\epsilon} &=& 0, \\
\Pi_3^{abc\lambda\tau\epsilon} &=& -g^3f^{abc}\frac{81}{54\pi^2} \epsilon^{\mu\lambda\tau\epsilon}b_\mu.
\end{eqnarray}
\end{subequations}
As observed above, despite the integrals being divergent by power counting, the tensors are finite, resulting in the sum
\begin{eqnarray}
\label{threemin1aa}
\Pi^{abc\lambda\tau\epsilon} &=& \Pi_1^{abc\lambda\tau\epsilon} +\Pi_2^{abc\lambda\tau\epsilon} +\Pi_3^{abc\lambda\tau\epsilon} \nonumber\\
&=& -g^3f^{abc}\frac{61}{27\pi^2} \epsilon^{\mu\lambda\tau\epsilon}b_\mu,
\end{eqnarray}
which has the same coefficient of $\Pi^{ab\lambda\tau}$ (\ref{twomin1aaa}), as expected for the gauge invariance. Then, we obtain the result perfectly replaying the non-Abelian CFJ term \cite{ColMac,ourYM,ptime}:
\begin{eqnarray}
{\cal L}_{CFJ}={\rm tr}\, \kappa_{\alpha}\epsilon^{\alpha\mu\nu\rho}(A_{\mu}\partial_{\nu}A_{\rho}-\frac{2}{3} ig A_{\mu}A_{\nu}A_{\rho}),  
\end{eqnarray}
where the vector $\kappa_{\alpha}$ is finite and, as we will see below, ambiguous. Explicitly, within the scheme we use here,
\begin{equation}
\kappa_{\alpha}=\frac{61g^2}{54\pi^2}b_{\alpha},
\end{equation}
i.e., the result is finite.

Let us now use an alternative prescription for calculation, namely,  the 't Hooft-Veltman regularization scheme \cite{tHoVel}. In this prescription, we split the $D$-dimensional $\gamma^\mu$ and $g^{\mu\nu}$ into $4$-dimensional and $(D-4)$-dimensional parts, i.e., $\gamma^\mu=\bar\gamma^\mu+\hat\gamma^\mu$ and $g^{\mu\nu}=\bar g^{\mu\nu}+\hat g^{\mu\nu}$, and consider the relations $\{\bar\gamma^\mu,\bar\gamma^\nu\}=2\bar g^{\mu\nu}$, $\{\hat\gamma^\mu,\hat\gamma^\nu\}=2\hat g^{\mu\nu}$, $\{\bar\gamma^\mu,\hat\gamma^\nu\}=0$, $\{\bar\gamma_\mu,\gamma_5\}=0$, and $[\hat\gamma_\mu,\gamma_5]=0$, with the contractions $\bar g_{\mu\nu}\bar g^{\mu\nu}=4$, $\hat g_{\mu\nu}\hat g^{\mu\nu}=D-4$ and $\bar g_{\mu\nu}\hat g^{\mu\nu}=0$. 

Thus, for the two-point function expressions (\ref{subtwomin1}), by taking into account these rules and performing the trace, we obtain
\begin{subequations}
\begin{eqnarray}
\label{twomin1a0}
\Pi_{1a}^{ab\lambda\tau} &=& \Pi_{2a}^{ab\lambda\tau} = \Pi_{2b}^{ab\lambda\tau} \nonumber\\
&=& \mu^{4-D}\int\frac{d^Dk}{(2\pi)^D}\frac{4ig^2 \delta^{ab}\epsilon^{\mu\nu\lambda\tau}b_\mu p_\nu}{(D-1)^2 D m^2} \left[\frac{2 (D-2) ((2 D-11) D+18)}{k^2-m^2}\right. \nonumber\\
&& +\frac{(D (((D-10) D+15) D+94)-256) m^2}{(k^2-m^2)^2} \nonumber\\ 
&& \left.+\frac{4 \left(-D^3+6 D^2+D-46\right) m^4}{(k^2-m^2)^3}\right].
\end{eqnarray}
\end{subequations}
The tensors yield the same result, which after the integration vanishes, i.e., we get
\begin{equation}
\label{tHV2}
\Pi_{1a}^{ab\lambda\tau}=\Pi_{2a}^{ab\lambda\tau}=\Pi_{2b}^{ab\lambda\tau}=0.
\end{equation}

Finally, for three-point function expressions (\ref{subthreemin}), we obtain
\begin{subequations}
\begin{eqnarray}
\label{threemin1b}
\Pi_1^{abc\lambda\tau\epsilon} &=& \Pi_2^{abc\lambda\tau\epsilon} \nonumber\\
&=& \mu^{4-D}\int\frac{d^Dk}{(2\pi)^D}\frac{4ig^2 \delta^{ab}\epsilon^{\mu\nu\lambda\tau}b_\mu p_\nu}{(D-1)^2 D m^2} \left[\frac{2 (D-2) ((2 D-11) D+18)}{k^2-m^2}\right. \nonumber\\
&& +\frac{(D (((D-10) D+15) D+94)-256) m^2}{(k^2-m^2)^2} \nonumber\\ 
&& \left.+\frac{4 \left(-D^3+6 D^2+D-46\right) m^4}{(k^2-m^2)^3}\right], \\
\Pi_3^{abc\lambda\tau\epsilon} &=& -\mu^{4-D}\int\frac{d^Dk}{(2\pi)^D}\frac{4ig^3 f^{abc}\epsilon^{\mu\lambda\tau\epsilon}b_\mu}{(D-1)^2 D m^2} \left[\frac{2 (D-6) (D-2)}{k^2-m^2}\right. \nonumber\\ 
&& +\frac{((D (D ((D-11) D+39)-15)-210) D+384) m^2}{(k^2-m^2)^2} \nonumber\\ 
&& \left.-\frac{4 (D (((D-7) D+11) D+29)-90) m^4}{(k^2-m^2)^3}\right].
\end{eqnarray}
\end{subequations}
Now, only two tensors have the same result, but after the integration, all results vanish, so we write
\begin{equation}
\Pi_1^{abc\lambda\tau\epsilon}=\Pi_2^{abc\lambda\tau\epsilon}=\Pi_3^{abc\lambda\tau\epsilon}=0,
\end{equation}
in which, together with (\ref{tHV2}), the gauge invariance is maintained again. Therefore, within this prescription, we have $\kappa_{\alpha}=0$, i.e., in this case, the CFJ term vanishes. 

Thus, we conclude that in the non-Abelian case, this term is ambiguous just as in the Abelian one \cite{prev}, which apparently means that ambiguity is an intrinsic property of the CFJ term independent of the theory where it arises. We note that the presence of the zero result for this term, as well as for other ambiguous terms arising in other theories, is rather natural. Some advantages of zero results for different ambiguous terms are discussed in \cite{Alt,Alt1}.

\section{Summary}

We succeeded to generate the non-Abelian CFJ term from coupling of the gauge field to the Rarita-Schwinger one.  Despite highly divergent loop integrals, the result turns out to be finite, both its quadratic and cubic parts. As it is typical for LV theories, the finiteness of the superficially divergent contribution signalizes about its ambiguity. This ambiguity is confirmed within our calculations, so that at least two different results for the CFJ coefficient $\kappa_{\mu}$ are possible. In total analogy with the usual QED, it is natural to expect that other values for $\kappa_{\mu}$ can be found as well.  Besides of this, we demonstrated explicitly that the result is compatible with the gauge symmetry reproducing the standard relations of coefficients accompanying quadratic and cubic terms. 

We found that within the 't Hooft-Veltman prescription, our result for the CFJ term vanishes. This is rather natural, taking into account that, as it was argued in \cite{JK}, for the usual LV QED with only $b_{\mu}$ LV parameter, the 't Hooft-Veltman prescription implies in a gauge invariance of the perturbative contribution to the Lagrangian (and not only the corresponding action), which requires the CFJ term to vanish (for the discussion of a relation between a generalization of the $\gamma_5$ matrix and the ambiguity of the CFJ term, see also \cite{Alts2004}). It is natural that the similar situation occurs within other schemes of generating the CFJ term, including in our theory.
We note that the ambiguity of a quantum correction is known to be related with anomalies, for the CFJ term -- with the axial anomaly \cite{JackAmb}. Therefore, it is natural to expect that the axial anomaly can arise in a theory involving gauge and RS fields (we note that some studies of anomalies in theories involving RS fields, but without Lorentz symmetry breaking, were performed earlier, see \cite{Adler1,Adler2}, so, studying, first, of a possible relation between these anomalies and CFJ term and, second, of a possible generalization of these anomalies to the LV case, is certainly an interesting problem).

A natural continuation of this study could consist in coupling of the Rarita-Schwinger field to the gravity, in a Lorentz-breaking matter, with generating the four-dimensional gravitational Chern-Simons term in a manner similar to \cite{ptime}, and calculation of the possible related gravitational anomalies. We expect to perform this study in a forthcoming paper.

{\bf Acknowledgments.}  This work was partially supported by Conselho Nacional de Desenvolvimento Cient\'\i fico e Tecnol\'ogico (CNPq). The work of A. Yu.\ P. has been partially supported by the CNPq project No. 301562/2019-9.

\end{document}